\documentclass[pra,twocolumn,showpacs]{revtex4}
\usepackage{epsfig,amsmath,amssymb,mathtools}
\usepackage[active]{srcltx}
\usepackage[hypertex,linkcolor=red]{hyperref}
\usepackage{yfonts}
\def\<{\langle}\def\>{\rangle}\def\set#1{{\sf #1}}
 \def\vec#1{\mathbf{#1}}
\newtheorem{lemma}{Lemma}

\def\Proof{\medskip\par\noindent{\bf Proof.
  }}\def\qed{$\,\blacksquare$\par}
\def\eg{e.~g. }\def\ie{i.~e. }
\def\N{\set{N}}
\def\jp#1{\set{J}^-_#1}\def\jf#1{\set{J}^+_#1}\def\j#1{\set{J}_#1}
\def\F{\set{L}}
\def\c#1{\set #1}
\def\cC{\c C}
\def\cO{\c O}
\def\R{\textfrak{R}}
\def\a{\alpha}
\def\b{\beta}
\begin{document}
\title{Space-time and special relativity from causal networks}
\author{Giacomo Mauro D'Ariano} \affiliation{QUIT Group, Dipartimento di
  Fisica ``A. Volta'', via Bassi 6, I-27100 Pavia, Italy and Istituto
  Nazionale di Fisica Teorica e Nucleare, Sezione di Pavia.}
\author{Alessandro Tosini} \affiliation{QUIT Group, Dipartimento di Fisica
  ``A. Volta'', via Bassi 6, I-27100 Pavia, Italy and Istituto
  Nazionale di Fisica Teorica e Nucleare, Sezione di Pavia.}
\begin{abstract}
  We show how the Minkowskian space-time emerges from a topologically homogeneous causal network,
  presenting a simple analytical derivation of the Lorentz transformations, with metric as pure
  event-counting. The derivation holds generally for $d=1$ space dimension, however, it can be
  extended to $d>1$ for special causal networks.
\end{abstract}
\pacs{03.30.+p, 04.20.Gz}
\maketitle  
Do events happen in space-time or is space-time that is made up of events? This question may be
considered a "which came first, the chicken or the egg?'' dilemma, but the answer may contain the
solution of the main problem of contemporary physics: the reconciliation of quantum theory (QT) with
general relativity (GR). Why? Because ``events'' are central to QT and ``space-time'' is central to
GR.  Therefore, the question practically means: which comes first, QT or GR?

In spite of the evidence of the first position---''events happen in space-time''---the second
standpoint---''space-time is made up of events''---is more concrete, if we believe \`a la Copenhagen that
whatever is not ``measured'' is only in our imagination: space-time too must be measured, and
measurements are always made-up of events. Thus QT comes first. How? Space-time emerges from the
tapestry of events that are connected by quantum interactions, as in a huge quantum computer: this
is the Wheeler's {\em It from bit} \cite{Wheeler}. For a theory of quantum gravity a variation of QT may
still be needed, such as a ``third-quantization'' of causal connections, allowing non
pre-established causal relations.  However, at least for the simplest case of Special Relativity
(SR) QT tout court should be sufficient. Ref. \cite{DAriano:QCFT} showed the mechanism with which
space-time emerges endowed with SR from a network of causally connected events, starting only from
the topology of the network, and getting the metric from pure event-counting. Ref.  \cite{Knuth}
later has shown how the Minkowski signature can be derived from the causal poset. Here we will
present a simple analytical derivation of the Lorentz transformations from a causal network (CN) in
1 space dimension: generalization to larger dimensions will be discussed at the end of the paper. As
we will see, the only thing that is needed in addition to causality is the topological homogeneity
of the CN, corresponding to the Galileo relativity principle.

The program of deriving the geometry of space-time from purely causal structure (causal sets) is not
new, and was initiated by Sorkin and collaborators more than two decades ago
\cite{Bombelli-Sorkin_(1987)}. In this publication and in following ones (see the review
\cite{Surya2008}) the possibility of recovering the main features of the space-time
manifold---topology, differentiable structure and the conformal metric---has been investigated,
starting from discrete sets of points endowed with a causal partial ordering. Since from the start,
causal sets were an independent research line in quantum gravity, since they naturally possess a
space-time discreteness and provide a history-space for a ``path integral'' formulation
\cite{Fotini2002,Henson2006}. They also fit perfectly the spirit of very recent works on operational
probabilistic theories \cite{Hardy,CDP}, and closely resemble Lamport's clock syncronization problem
in distributed computation \cite{Lamport}.

We now introduce the main notion of causal network (CN) as a partially ordered set of events with
the partial order representing the causal relation between two events. As mentioned, the aim is to
have the space-time endowed with SR emerging from the network of events, thinking to them not as
``happening in space-time'', but as making up space-time themselves. Thus the notions of {\em event}
and {\em causal relation} have to be considered as primitive, similarly to those of ``point'' and
``line'' in geometry (for their meaning in an operational framework and in QT, see Refs.
\cite{CUP,CDP}).  In synthesis, the CN represents the most general structure of  ``information
processing''.

\begin{figure}[h]
  \includegraphics[width=.225\textwidth]{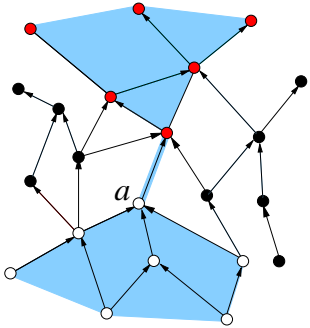}\quad\;
  \caption{Causal network: illustration of the set of past/future
    light-cone of event $a$.}\label{fig:DAG}
\end{figure}

A {\em causal set} is a set $\N$ of elements called {\em events} $a,b,c\ldots,\in\N$ equipped with a
partial order relation $\preceq$ which is: (1) {\em Reflexive:} $\forall a\in\N$ we have $a\preceq
a$; (2) {\em Antisymmetric:} $\forall a,b\in\N$, we have $a\preceq b\preceq a\Rightarrow a=b$; (3)
{\em Transitive:} $\forall a,b,c\in\N$, $a\preceq b\preceq c\Rightarrow a\preceq c$; (4) {\em
  Locally finite:} $\forall a,c\in\cC$, $|\{b\in\N:a\preceq b\preceq c\}|<\infty$, where $|\set{S}|$
denotes the cardinality of the set $\set{S}$. In the following we will also write $a\prec b$ to
state that $a\preceq b$ with $a\neq b$. A causal set is represented by a \emph{graph} with points
being the events and the edges drawn between any two points $a$ and $b$ for which $a\preceq b$---\ie
that are causal connected, as in Fig.  \ref{fig:DAG}. What we call a {\em causal network} (CN) is a
causal set unbounded in all directions. In order to satisfy transitivity, the CN is a {\em
  directed acyclic graph}, \ie loops are forbidden (arrows on edges are usually not drawn by
orienting the graph \eg from the bottom to the top).

Causality of the network naturally suggests the notion of {\em light-cone} $\j{a}$ of an event
$a\in\N$, along with those of {\em past/future light-cone} $\jp{a}$/$\jf{a}$, respectively (see Fig.
\ref{fig:DAG})
\begin{equation}
    \jp{a}:=\{b\in\N:b\preceq a\},\quad
   \jf{a}:=\{b\in\N:a\preceq b\},
\end{equation}
and $\j{a}:=\jp{a}\cup\jf{a}$.  Accordingly, one has that $a\preceq b$ is equivalent to $a\in\jp{b}$
and to $b\in\jf{a}$. We will call {\em independent} or {\em space-like} two events $a,b\in\N$ that
are not causally related---namely $a\not\in\j{b}$ (or $b\not\in\j{a}$)---and {\em causally
  dependent} or {\em time-like} otherwise, namely when $ a\in\j{b}$ (or $b\in\j{a}$).  We call a CN
{\em connected} if for every $a,b\in\N$ there exists $c\in\j{a}\cap\j{b}$, corresponding to the
intuitive notion of connectedness.
Two events that are not space-like are connected by at least a causal chain, \eg\ $a\preceq b$ are
connected by the causal chain $\cC(a,b)$ given by $\cC(a,b):=\{c_i\}_{i=1}^N$, with $a\equiv
c_1\prec c_2\prec\ldots c_N\equiv b$. Being the equivalent of a {\em world-line}, the causal chain
plays also the role of an {\em observer}. It is convenient to orient the chain, generalizing its
definition to include the case $b\preceq a$, writing $\cC(a,b)$ for $\cC(a,b):=\{c_i\}_{i=1}^N$,
with $b\equiv c_1\prec c_2\prec\ldots c_N\equiv a$. The verse of the chain is taken into account by
a signed cardinality $|\cC(a,b)|_\pm:=\sigma|\cC(a,b)|$ with $\sigma=+$ for $a\prec b$, and
$\sigma=-$ for $b\prec a$.

\begin{figure}[h]
  \includegraphics[width=.44\textwidth]{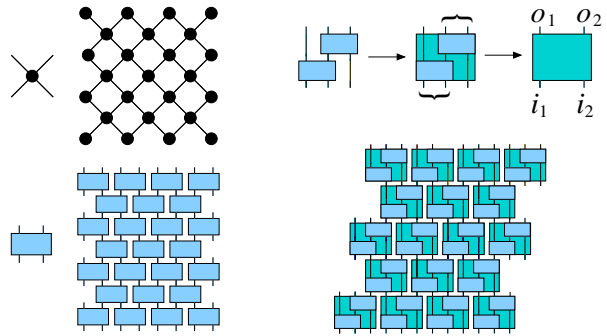}
  \caption{Left: homogeneous causal network and equivalent representation as a
 quantum circuit. Right: example of coarse-graining preserving the homogeneity of the network.
 \label{fig:causal_net}}
\end{figure}

In order to derive SR from the CN, we need the equivalent of the Galileo principle \cite{AJP},
namely the invariance of the physical law with the reference system.  Within a single frame the
Galileo principle is just uniformity of space and time. In the present purely topological context,
this translates to the topological homogeneity of the CN, the physical law being the causal
connection-rule of the network, \ie the tile of the causal pattern. At this point, we need to make
more specific the notion of CN, introducing different types of links, \eg in Fig.
\ref{fig:causal_net} we have two generally different kinds of input links---the left and the right
ones---for each node. It is now convenient to label links with letters.  We then consider the input
and the output sets $\set{l}_{in}(a)=\{i_1(a),i_2(a),\ldots i_K(a)\}$ and
$\set{l}_{out}(a)=\{o_1(a),o_2(a),\ldots o_H(a)\}$ of links of an event. We now say that a CN is
{\em topologically homogeneous} if for each couple of events $a,b\in\N$ one has the isomorphism
$i_j(a)=i_j(b)$ and $o_j(a)=o_j(b)$ for $j=1,\ldots H=K$. An example of homogeneous CN is given in
Fig. \ref{fig:causal_net}.  There is no loss of generality in considering only homogeneous CN with
$H=K$ and with all events isomorphic: in fact, one can always reach this situation, by grouping
connected events into single ones, \ie by {\em event coarse-graining}  (see e.~g. Fig \ref{fig:causal_net}).

In a homogeneous causal network we can also easily see how causality is sufficient to guarantee a
maximum speed of ``information flow''. Such speed is just ``one-event per step'', corresponding to a
line at $45^o$ in Fig. \ref{fig:causal_net} (to connect events along a line making an angle $< 45^o$
with the horizontal, one needs to follow some causal connections in the backward direction from the
output to the input).

We will now introduce the notion of simultaneity in relation to an observer. The observer is just a
causal chain (conveniently taken as unbounded). We label the events of the chain with relative
numbers, choosing an event for the zero. Hence, an observer will be denoted as
$\cO_a=\{o_i\}_{i\in\mathbb Z}$, with $o_i\preceq o_{i+1}$ $\forall i\in\mathbb Z$, and with $a=o_0$
representing the origin. The index $i\in\mathbb Z$ plays the role of the observer's proper time.
Thanks to the topological homogeneity, we can translate the observer $\cO_a$ to any event $a'\in\N$.
We will denote by $\cO$ the equivalence class of all observers translated over all events of the CN.
We will also denote by $\cO_a(b,c)$ the causal chain $\cC(b,c)\subset\cO_a$.  We now define
simultaneity of events $a$ and $b$---denoted as $a\sim_\cO b$---as follows
\begin{equation}
  a\sim_\cO b\Leftarrow \inf_{{b^*}\in \jf{b}}|\cO_a
  (a,b^*)|_\pm= \inf_{{a^*}\in \jf{a}}|\cO_b (b,a^*)|_\pm.
\end{equation}
Depending on the shape of the observer chain, one may have situations in which there are no
synchronous events. However, it is easy to see that for an observer that is topologically
homogeneous (\ie periodic) there always exist infinitely many simultaneous events.  Moreover, modulo
event coarse-graining, without loss of generality we can restrict only to observers with a zig-zag
with a single period, with $\alpha\geq 1$ steps to the right and $\beta\geq 1$ steps to the left (we
will call them {\em simply periodic}). Each ziz-zag is the equivalent of a tic-tac of an Einstein
clock made with light bouncing between two mirrors. All events on the same mirror lay on a line, and
for such events there always exist (infinitely many) synchronous events.

The given notion of simultaneity allows us to associate each observer with a {\em foliation} of the
CN. For each event $o_i\in\cO_a$ there is a {\em leaf}~ $\F_i(\cO_a)$, which is the set of events
simultaneous to $o_i$ with respect to the observer $\cO_a$, namely
\begin{equation} 
  \F_i(\cO_a):=\{b\in\N:b\sim_\cO o_i\}.
\end{equation}
The collection of all leaves for all the events in $\cO_a$ is the {\em foliation} $\F(\cO_a)$ of
$\N$ associated to the observer $\cO_a$
\begin{equation}
  \F(\cO_a):=\{\F_i(\cO_a),\forall i\in{\mathbb Z}\}.
\end{equation}
The foliation has an ``origin'' $a$ defined by the observer $\cO_a$. Homogeneity of foliations
follows from that of the observer. Notice that a foliation does not generally contain all the events
of the CN (it certainly does for $\alpha=\beta=1$): this fact is related to the sparsness issue
raised in Ref. \cite{Dowker} for Lorentz-transformed regular lattices of points.

For a given foliation $\F(\cO_a)$ we can now define a pair of coordinates $\vec z(b)$ for any event
$b\in \F(\cO_a)$ via the map
\begin{equation}\label{e:cmap}
\begin{split}
 &K_{\cO_a}:\N\to\mathbb Z^2,\quad b\mapsto K_{\cO_a}(b):=\vec z(b)=
\begin{bmatrix}z_1(b)\\z_2(b)\end{bmatrix},\\
  z_1(b&)\coloneqq\inf_{b^*\in\jf{b}}|\cO_a (a,b^*)|_\pm,\quad
  z_2(b)\coloneqq\inf_{a^*\in\jf{a}}|\cO_b(b,a^*)|_\pm.
\end{split}
\end{equation}
\begin{figure}[h]\label{fig:coord_map}
  \includegraphics[width=.35\textwidth]{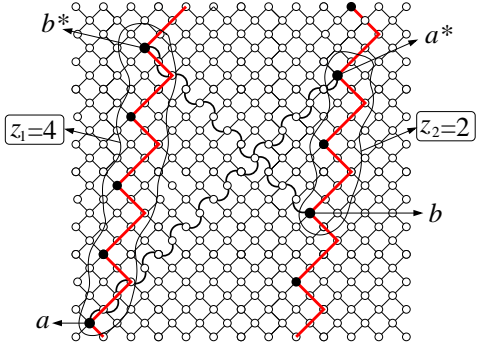}
  \caption{Illustration of the coordinate map in Eq. (\ref{e:cmap})
    (the observer has $\alpha=3$ and $\beta=2$).}
\end{figure}
Thus, to each observer $\cO_a$ it corresponds a coordinate map, and this is what is commonly called
a {\em reference frame}---shortly {\em frame}.  The coordinates $z_1$ and $z_2$ do not have an
immediate meaning, but get an simple interpretation thanks to the following Lemma.
\begin{lemma}\label{lemz} An event $b\in\F(\cO_a)$ belongs to the $t$-th leaf
  $\F_t(\cO_a)$ for
  $t=(z_1-z_2)/2$, and the number of events on such leaf between $b$
  and $\cO_a$ is given by $s=(z_1+z_2)/2$.
\end{lemma}
\Proof There exists $t\in\mathbb Z$ such that $o_t$ is simultaneous to $b$. By definition one has
$b\in\F_t(\cO_a)$, and
\begin{equation}\label{e:A}
  \inf_{b^*\in\jf{b}}|\cO_{o_t} (o_t,b^*)|_\pm=
  \inf_{{{o_t}^*}\in\{jf{o_t}}|\cO_b (b,{o_t}^*)|_\pm.
\end{equation}
One has
\begin{equation}\label{e:B}
  z_1(b)=t+\inf_{b^*\in\jf{b}}|\cO_{o_t} (o_t,b^*)|_\pm,
\end{equation}
whereas
\begin{equation}
  z_2(b)=\inf_{o_t^*\in\jf{{o_t}}}\inf_{a^*\in\jf{a}}\Big(
  |\cO_b (b,o_t^*)|_\pm+|\cO_b (o_t^*,a^*)|_\pm\Big).
\end{equation}
Topological homogeneity implies that
\begin{equation}\label{e:C}
  z_2(b)=\inf_{{{o_t}^*}\in\jf{{o_t}}}|\cO_b (b,o_t^*)|_\pm-t.
\end{equation} 
Using the simultaneity condition in Eq. (\ref{e:A}) we can combine Eqs.  (\ref{e:B}) and (\ref{e:C})
to get $t=\frac{1}{2}(z_1-z_2)$.\qed

According to the last Lemma the coordinates
\begin{equation}\label{e:stmap}
  \begin{bmatrix}t(b)\\s(b)\end{bmatrix}:=2^{\frac{1}{2}}\vec U(\pi/4)
  \begin{bmatrix}z(b)\\z(b)\end{bmatrix},
\end{equation}
where $\vec U(\theta)$ is the $\theta$-rotation matrix, is interpreted as the space-time
coordinates of the event $b$ in the frame $\F(\cO_a)$.
 
\paragraph*{Frames in standard configuration (boosted).}
Consider now two observers $\cO_a^1=\{o^1_i\}$ and $\cO_a^2=\{o^2_j\}$ sharing the same origin
(homogeneity guarantees the existence of observers sharing the origin).  We will shortly denote the
two frames as $\R^1$ and $\R^2$, and the corresponding coordinate maps as $K^1$ and $K^2$.  We will
say that the two frames $\R^1$ and $\R^2$ are in {\em standard configuration} if there exist
positive $\a^{12},\b^{12}$, such that $\forall i\in\mathbb Z$
\begin{equation}\label{e:obcoor}
  \quad  K^1(o_i^2)=\vec D^{12}K^2(o_i^2),\:\vec D^{12}:=
  \operatorname{diag}(\a^{12},\b^{12}). 
\end{equation}
It turns out that having chosen only simply periodic observers, one
has $\alpha^{ij}=\alpha^j/\alpha^i \in\mathbb Z^+$,
$\beta^{ij}=\beta^j/\beta^i\in\mathbb Z^+$.

\begin{figure}[h]
  \includegraphics[width=.40\textwidth]{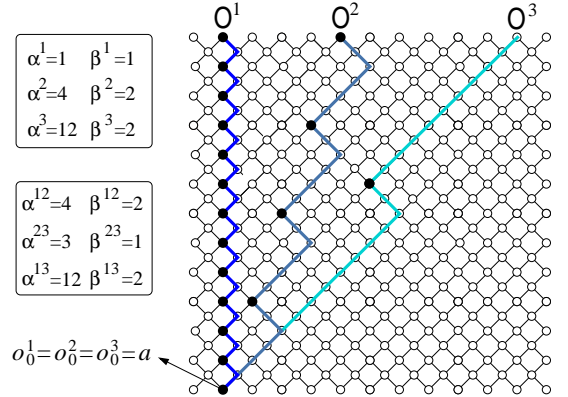}
  \caption{Example of three observers related as in
    Eq. (\ref{e:obcoor}) and then generating reference frames in
    standard configuration.\label{fig:std_obs}}
\end{figure}

Examples of observers corresponding to frames in standard configuration are shown in Fig.
\ref{fig:std_obs}. Clearly different frames correspond to generally different sets of events, and
what follows applies to the events in their intersection: thus, again, the transformation includes
an implicit event coarse-graining (see e.~g. Fig \ref{fig:causal_net}).
We now see how it is possible to define a relative velocity between two frames in standard
configuration. It is readily seen that $K^2(o^2_n)=(n,-n)$, whence
$K^1(o_l^2)=(l\a^{12},-l\b^{12})$. We can now define the relative velocity between $\R^1$ and $\R^2$
as the quotient between the space and time coordinates of the observer $\cO^2$ with respect to 
observer $\cO^1$, namely, by Lemma \ref{lemz}
\begin{equation}\label{e:velocity}
  v^{12}=\frac{n\a^{12}-n\b^{12}}{n\a^{12}+n\b^{12}}=\frac{\a^{12}-\b^{12}}{\a^{12}+\b^{12}}.
\end{equation}
Of course one has $K^2(o_i^2)=\vec D^{21}K^1(o_i^2)$ $\forall
i\in\mathbb Z$, with $\vec D^{21}=\vec
{D^{12}}^{-1}=\text{diag}(1/{\a^{12}},1/{\b^{12}})$, whence upon
rewriting Eq. (\ref{e:velocity}) for $v^{21}$ one obtains
$v^{21}=-v^{12}$.

\smallskip\paragraph*{Velocity-composition rule.}
\par Consider three frames $\R^1$, $\R^2$, $\R^3$ in pairwise standard
relation, associated to observers $\cO^1$, $\cO^2$, $\cO^3$ sharing
the origin $a$, corresponding to the coordinate maps $K^1,K^2, K^3$
(see for example the situation illustrated in
Fig. \ref{fig:std_obs}). Let $\vec
D^{12}=\text{diag}(\a^{12},\b^{12})$ and $\vec
D^{23}=\text{diag}(\a^{23},\b^{23})$ be the matrixes relating
respectively the coordinates of the second observer with respect to
the first one and the coordinates of the third observer with respect
to the second one, according to
\begin{equation}
  K^1(o_i^1)=\vec D^{12}K^2(o_i^2),\quad
  K^2(o_j^{3})=\vec D^{23} K^3(o_j^3).
\end{equation}
We are interested in the relation between the coordinates of frame
$\R^3$ with respect to frame $\R^1$. This is given by
\begin{equation}
  K^1(o_j^3)=\vec D^{13}K^3(o_j^3),
\end{equation}
with matrix $
  \vec D^{13}=\vec D^{12}\vec D^{23}=
  \operatorname{diag}(\a^{12}\a^{23},\b^{12}\b^{23})$.
From Eq. (\ref{e:velocity}) it immediately follows that 
\begin{equation}\label{e:network_vel_comp}
  v^{13}=\frac{\a^{12}\a^{23}-\b^{12}\b^{23}}
  {\a^{12}\a^{23}+\b^{12}\b^{23}},
\end{equation}
which by simple algebraic manipulations gives
\begin{equation}
  v^{13}=\frac{\left(\frac{\a^{12}-\b^{12}}{\a^{12}+\b^{12}}\right)+
    \left(\frac{\a^{23}-\b^{23}}{\a^{23}+\b^{23}}\right)}
  {1+\left(\frac{\a^{12}-\b^{12}}{\a^{12}+\b^{12}}\right)
    \left(\frac{\a^{23}-\b^{23}}{\a^{23}+\b^{23}}\right)}=
  \frac{v_{12}+v_{23}}{1+v_{12}v_{23}},
\end{equation}
namely the velocity composition rule of special relativity.  \smallskip
\paragraph*{Lorentz transformations.}
Again using Lemma \ref{lemz} we can derive the space-time coordinate
transformations between the two frames $\R^1$ and $\R^2$ in standard
relation. Using the topological homogeneity of $\N$ it follows that
Eq. (\ref{e:obcoor}) holds for any event $b\in\R^1\cap\R^2$. One has
$z_1^1=\a^{12}z_1^2$ and $z_2^1=\b^{12}z_2^2$, and after easy
manipulations we get
\begin{equation}\label{e:space-time}
  \frac{z_1^1\pm z_2^1}{2}=\frac{\a^{12}+\b^{12}}{2} \left[\frac{z_1^2\pm
      z_2^2}{2}+\left(\frac{\a^{12}-\b^{12}}{\a^{12}+\b^{12}}\right)
    \frac{z_1^2\mp z_2^2}{2} \right], 
\end{equation}
where we can easily identify the space-time coordinates of the event
in the two frames and their relative velocity, in terms of which
Eqs. (\ref{e:space-time}) become
\begin{equation}\label{ttrans}
\begin{split}
    t^1&=\tfrac{1}{2}(\a^{12}+\b^{12}) \left( t^2+v^{12}s^2
    \right),\\
    s^1&=\tfrac{1}{2}(\a^{12}+\b^{12}) \left( s^2+v^{12}t^2 \right).
\end{split}
\end{equation}
Using the simple relation
\begin{equation}\label{e:gamma}
  \tfrac{1}{2}(\a^{12}+\b^{12})=\frac{\chi_{12}}{\sqrt{1-(v^{12})^2}},\quad \chi_{12}:=
  \sqrt{\a^{12}\b^{12}},
\end{equation}
we obtain the identities
\begin{equation}
 t^1=\chi_{12}\frac{t^2+v^{12}s^2}{\sqrt{1-(v^{12})^2}},\qquad
 s^1=\chi_{12}\frac{s^2+v^{12}t^2}{\sqrt{1-(v^{12})^2}},
\end{equation}
which differ from the Lorentz transformations only by the
multiplicative factor $\chi_{12}$.  The factor $\chi_{12}$ can be
removed by rescaling the coordinate map in Eq.  (\ref{e:stmap}) using
the factor $(2\alpha\beta)^{\frac{1}{2}}$ in place of
$2^{\frac{1}{2}}$, with the constants $\alpha$ and $\beta$ of the
observer.  The relative velocity between two frames $\R^1$ and $\R^2$
does not change in this representation because the common factor
simplifies in Eq.  (\ref{e:velocity}). Consequently also the
velocity-composition rule is left unchanged. A multiplicative factor
$\sqrt{\frac{\alpha^{1}\beta^{1}}{\alpha^{2}\beta^{2}}}=\chi_{12}^{-1}$
now shows up after the factor $1/2$ in both transformations
(\ref{ttrans}), and, using relation (\ref{e:gamma}) we get the usual
Lorentz transformations.

We emphasize that the whole procedure for defining the space-time
coordinates is made only with event-counting on the CN.  For each
transformation a corresponding coarse-graining (of the starting or the
ending foliation) seems essential (corresponding to the usual
rescaling in the Minkowski space, due to reciprocity between the
observers). Finally, it is clear that our derivation could be extended
to $d>1$ space dimensions, for CN that are embeddable in $d+1$
dimensions, with leaves that can be embedded in $d$ dimensions, \eg
for a $d+1$-dimensional diamond lattice.

\end{document}